\documentclass[aps,prl,twocolumn,showpacs]{revtex4}
\usepackage{graphicx}
\newcommand{\create}[2]{\hat{\psi}^{\dagger}_{#1}({\bf #2})}
\newcommand{\destroy}[2]{\hat{\psi}^{\phantom\dagger}_{#1}({\bf #2})}
\newcommand{\screate}[2]{{\varphi}^{*}_{#1}({\bf #2})}
\newcommand{\sdestroy}[2]{{\varphi}^{\phantom *}_{#1}({\bf #2})}
\newcommand{\akcreate}[1]{\hat a^{\dagger}_{{\bf #1}}}
\newcommand{\akdestroy}[1]{\hat a^{\phantom\dagger}_{{\bf #1}}}

\begin{document}
 

\title{Resonant formation of strongly correlated paired states in rotating Bose gases}
 
\author{S. G. Bhongale, J. N. Milstein, and M. J. Holland}
 
\affiliation{JILA, University of Colorado and National Institute of
  Standards and Technology, Boulder, Colorado 80309-0440}
 

\begin{abstract}
We propose increasing the fractional quantum Hall gap of a rapidly
rotating Bose gas by increasing the interatomic interactions via a
Feshbach resonance. The generation of molecules by the resonance
causes pair correlations to grow throughout the system effecting the
ground state.  By an extension of the usual Chern-Simons theory, built
of composite atoms and molecules, we are able to account for these
resonance effects.  We find that the resulting ground state evolves
from a Laughlin wavefunction to a unique paired wavefunction as one
approaches the resonance.
\end{abstract}

\maketitle

\begin{figure}
  \includegraphics[scale=.35]{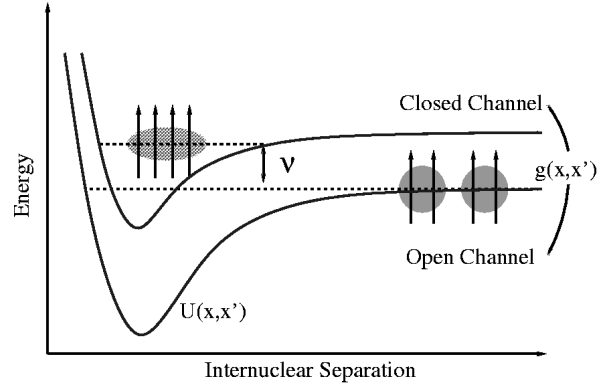}
\caption{The Feshbach resonance pairing mechanism is illustrated by
the above Born-Oppenheimer curves.  Pairs of composite atoms composed
of single atoms and an associated number of quanta of angular momentum
(represented by the arrows) approach each other within an open channel
potential of background value $U({\bf x},{\bf x'})$. They may form a
composite molecule due to the presence of a closed channel bound
state, at a detuning $\nu$ from the scattering continuum, which is
coupled to the open channel with strength $g({\bf x},{\bf x'})$. }
\label{CompositeRes}
\end{figure}

Recently, there has been considerable interest from both the
experimental and theoretical perspective in the behavior of rotating
Bose gases confined to an effective two-dimensional
space~\cite{engels,shaeer,ho,wilkin,paredes,fisher}.  One goal has
been to understand and to create strongly correlated states, such as
the Laughlin state generated by the fractional quantum Hall effect
(FQHE), within ultracold atomic gases. A major challenge to realizing
such a state is the need to reach extremely low temperatures in order
to resolve the lowest Laughlin state from the first excited state.
The size of this energy gap, however, is directly related to the
strength of interatomic interactions, so by increasing the
interactions it would seem natural that one could increase the gap,
making the system more accessible to experiment.\\ \indent Atomic
systems can now be created which allow the microscopic interactions to be
dynamically tuned~\cite{cornish,loftus}.  Feshbach resonances have
proven very successful at doing this, allowing one to tune the interactions by adjusting the resonant detuning $\nu$ (for example, by varying a magnetic field),  and would seem an excellent tool
for increasing the gap within a FQHE system.  To account for the full
effects of the resonance, however, we cannot simply scale the
mean-field energy but must incorporate the entire resonant structure
into our model.  This means that we must include the process of
molecular formation generated by the introduction of a bound state
within the open channel of scattering
states~\cite{holland,timmermans,kohler}.\\ \indent By introducing a
bound state, however, we not only modify the relative interaction
strength, but we also introduce a physical mechanism for generating
pair correlations between particles (see Fig.~\ref{CompositeRes}). In
the context of two-dimensional condensed matter systems, such a
mechanism, although arising from a very different
source~\cite{haldane}, can have a significant effect on the ground
state properties. Before we can study the resonant behavior of the
gap, we must first understand the effect of resonant interactions on
the ground state wavefunction.  We will find that as we approach the
Feshbach resonance the Laughlin state transforms into a unique,
strongly correlated state.\\
\indent We begin by writing down an effective Hamiltonian, in
second quantized form, for a resonant gas of Bosons of mass $m$
rotating in two dimensions with stirring frequency $\Omega$
approaching the trapping frequency 
\mbox{$\omega$, i.e. $\omega-\Omega \rightarrow 0^+$}:

\begin{eqnarray}
\!\!\hat{H}\!\! &=&\!\!\int\!\!d^2x\
\create{a}{x}\left[\frac{-1}{2m}(\nabla-i {\bf A} ({\bf
x}))^2\right]\destroy{a}{x}\label{hamiltonian} \\
\!\!&+&\!\!\frac{1}{2}\int\!\! d^2x'\!\!\int\!\!  d^2x\
\create{a}{x}\create{a}{x'}U({\bf x},{\bf
x'})\destroy{a}{x'}\destroy{a}{x}\nonumber\\ \!\!&+&\!\!\int\!\! d^2x\
\create{m}{x}\left[\frac{-1}{4m}(\nabla-2 i {\bf A} ({\bf
x}))^2+\nu\right]\destroy{m}{x}\nonumber\\
\!\!&+&\!\!\frac{1}{2}\int\!\!d^2x'\!\!\int\!\!d^2x\!\!\left[\create{m}{\frac{x+x'}{\text
2}}g({\bf x},{\bf x'})\destroy{a}{x}\destroy{a}{x'}+{\rm
H.c.}\right]\nonumber \!.
\end{eqnarray}

\noindent Here $\create{a,m}{x},\destroy{a,m}{x}$ are the creation and
destruction operators for atoms and molecules which satisfy the
commutation relations $[\destroy{1}{x},\create{2}{x'}]=\delta^{(3)}({\bf
x},{\bf x'})\delta_{1,2}$, where $1,2\!\in\!\{a,m\}$. We define the two-dimensional vector
potential ${\bf A} ({\bf x})=(m\omega y,-m\omega x)$, $U({\bf x},{\bf
x'})$ is the two-particle background scattering potential, $g({\bf
x},{\bf x'})$ is the resonant coupling between the open and closed
channel potentials, and $\nu$ is the detuning of the open channel
continuum from the level of the bound state in the closed channel.\\
\indent Recently, it has been noted that there is a direct mapping
between the Hamiltonian for a rotating two-dimensional gas, the first
two terms of Eq.~(\ref{hamiltonian}), and that of the
FQHE~\cite{wilkin}. The analog of the magnetic field is realized by
the angular rotation and the Coulomb interaction is replaced by the
two-particle scattering.  Therefore, if we were to neglect resonant
effects we would expect the many-body ground state to be described by
the Laughlin wavefunction~\cite{laughlin}

\begin{equation}
\Psi_L({\bf x}_1, {\bf x}_2,...{\bf x}_N)=\prod_{i<j}(z_i-z_j)^2\prod_k
\exp\left(-{|z_k|^2}/2\right),\label{laughlin}
\end{equation} 
where the products run over the indices $i,j,k=(1,2,...N)$ at position
$z=x+iy$ for $N$ particles.  We will see that the inclusion of the resonant terms in
Eq.~(\ref{hamiltonian}) can significantly modify the form of
Eq.~(\ref{laughlin}) due to the growth of two-particle correlations.
\\ \indent We approach this problem by an extension of the
Chern-Simons theory \cite{zhang} which allows us to develop a
mean-field theory for the rotating system that has removed the
complications of the associated rotation. This is done by constructing
a composite particle composed of the original particle and an
artificially attached number of flux quanta. The composite particle is
designed so that the attached flux quanta cancel the total rotation of
the original system leaving a system of non-rotating, interacting
composite particles.  For the resonant system, the composite particles
can formally be obtained by the following transformation:

\begin{eqnarray}
\hspace{-5mm}\hat{\varphi}_{a,m}({\bf x})=\exp\bigg[\!&-&\!q\int
d^2x'\theta({\bf x}-{\bf
x'})\nonumber\\\!&\times&\![\rho_a({\bf x'})+2\rho_m({\bf x'})]\bigg]\destroy{a,m}{x},
\end{eqnarray}
where $\rho_a({\bf x'})$ and $\rho_m({\bf x'})$ are the atomic and
molecular spatial densities, respectively, and $\theta({\bf x}-{\bf
x'})$ is the topological phase. To guarantee that the resulting
composite particle is a Boson, with no loss of generality we set
$q=2$.\\ \indent In the composite picture, we replace the $\hat\psi$
operators in Eq.~(\ref{hamiltonian}) with the corresponding composite
operators $\hat\varphi$ and introduce the statistical Chern-Simons
field ${\bf a}({\bf x})$ through the gauge transformation

\begin{eqnarray}
{\bf A}({\bf x})\rightarrow{\bf A}({\bf x})+{\bf a}({\bf x})
\end{eqnarray} 
These modifications generate the Hamiltonian formulation of our
composite atom/molecule theory.  The composite picture can be shown to
be equivalent to the single particle picture of
Eq.~(\ref{hamiltonian}).\\ \indent We will now shift to a functional
representation of the composite atom/molecule system to clarify the
resonant modifications of the Chern-Simons theory and then return to
the Hamiltonian formulation to derive the ground state wavefunction.
Defining the action within one temporal and two spatial dimensions

\begin{equation}
\!\!\!S=\int\!\!d^3x\sum_{\sigma=a,m}\screate{\sigma}{x}i\partial_0\sdestroy{\sigma}{x}-\int\!\!dx_0\hat{H},
\end{equation}
we generate a Chern-Simons term which couples to the statistical
vector field $a_\mu({\bf x})$

\begin{equation}
S_{CS}=-\int\!\!d^3x \frac{1}{8 \pi}\epsilon^{\mu \nu
\lambda} a_\mu({\bf x})\partial_\nu a_\lambda({\bf x}),
\end{equation}
where the indices of $\mu,\nu,$ and $\lambda$ run over the three dimensions $(0,1,2)$ and the summation
convention over repeated indices is invoked. We have also introduced the antisymmetric tensor $\epsilon^{012}=1$.  To simplify the
following calculations we assume contact interactions of the form
$U({\bf x},{\bf x'})=U\delta({\bf x},{\bf x'})$ and $g({\bf x},{\bf
x'})=g\delta({\bf x},{\bf x'})$. Any complications arising from this
replacement of the true potentials with contact potentials should be
remedied as explained in Ref.~\cite{kokonorm}.
We next perform the lowest order variation of the action.\\ \indent
Varying with respect to the zeroth component of the gauge field,
$\partial S/\partial a_0=0$, reproduces the Chern-Simons condition

\begin{equation}
\nabla\times {\bf a}({\bf
x})\big|_z=-4\pi\left(|\sdestroy{a}{x}|^2+2|\sdestroy{m}{x}|^2\right)\label{gauss}.
\end{equation}
Equation~(\ref{gauss}) is a statement of Gauss's law for the
statistical gauge field associating an even number of rotational flux
quanta with each particle.  This relation is simply a restatement of
our choice of quasiparticle.\\ \indent Since we will be interested in
the ground state properties of the atom/molecule system, let us assume
that the fields $\varphi_a({\bf x})$ and $\varphi_m({\bf x})$ are
uniform.  By minimizing the action with respect to the atomic and
molecular fields, i.e. $\partial S/\partial \varphi_a=0$, $\partial
S/\partial \varphi_m=0$, we generate the following constraint equation
for the molecules:

\begin{equation}
\varphi_m=\frac{g\varphi_a^2}{2[\nu+|{\bf
A} +{\bf a}|^2/m]}\label{molconst}.
\end{equation}
Equation~(\ref{molconst}) allows us to eliminate the molecular field
from the theory and arrive at a self-consistent relationship for the
gauge field

\begin{equation}
\left|{\bf A}+{\bf a}\right|^2 =\left(U+\frac{g^2}{4(\nu+|{\bf A}+{\bf a}|^2/m)}\right)
2m|\varphi_a|^2.\label{gaugeconst}
\end{equation}
Equation~(\ref{gaugeconst}) is the usual result relating the gauge field
to the background density only now it is dependent upon the detuning
from the resonance.

\indent We now switch back to the Hamiltonian form of our theory to
derive the ground state wavefunction.  After Fourier transforming the
composite form of Eq.~(\ref{hamiltonian}) by substitution of the field
operators $\hat\varphi_a({\bf x})=\sum_{\bf k} \akdestroy{k} e^{i {\bf k\cdot
x}}$ and $\hat\varphi_m({\bf x})=\sum_{\bf k} \hat b_{\bf k} e^{i {\bf k\cdot
x}}$ , we follow the usual Hartree-Fock-Bogolubov (HFB) approach to
construct a quadratic Hamiltonian which accounts for the lowest order
pairing.  As before, we assume contact interactions and now make the
additional assumption that we may neglect the excited modes of the
molecular field keeping only the lowest condensed mode $\hat b_{\bf
k}=b_0$.  The resulting Hamiltonian for the composite system can be
written in the form

\begin{equation}
H=H^0+\sum_{{\bf k}\neq 0} A_{\bf k}^\dagger M_{\bf k} A_{\bf k}.\label{dens}
\end{equation}
$H^0$ is composed of all terms of less then quadratic order in the
operator $\akdestroy{k}$, we define a column vector $A_{\bf k}=(\akdestroy{k}, \akcreate{-k})$,
and $M_{\bf k}$ is the self-energy matrix. For our purposes we need only
concern ourselves with the structure of the second term in
Eq.~(\ref{dens}).  Here the self-energy matrix is expressed as

\begin{equation}
M_{\bf k}=\left(\begin{array}{cc}E_{\bf k} &\Delta \\ \Delta^* &
E_{{\bf -k}}\end{array}\right)
\end{equation}
with the additional definitions for the diagonal and off-diagonal terms

\begin{eqnarray}
E_{\bf k} &=& E_{\bf k}^0+U(|\varphi_a|^2+n)\label{diag}\\ 
\Delta &=&
U(\varphi_a^2+p)+g\varphi_m \label{offdiag}.
\end{eqnarray}
Equations~(\ref{diag}) and ~(\ref{offdiag}) are expressed in terms of the
pairing-field $p=\sum_{\bf k'}\big<\akdestroy{k'} \akdestroy{-k'}\big>$,
the normal-field $n=\sum_{\bf k'}\big< \akcreate{k'} \akdestroy{k'}\big>$,
and $E_{\bf k}^0$ is the effective kinetic term which contains the
contribution from the gauge field $A({\bf x})$.
\\\indent Eq.~(\ref{dens}) can be rewritten in terms of the quasiparticles

\begin{eqnarray}
\hspace{-1cm}\hat \alpha_{\bf k} \!\!\!&=&\!\!\!\frac{1}{\sqrt{(E_{\bf
k}+\omega_{\bf k})^2+|\Delta|^2}}\left((E_{\bf k}+\omega_{\bf k})
\akdestroy{k}+\Delta \akcreate{-k}\right)\label{quasipart1}\\
\hspace{-1cm}\hat\alpha_{\bf -k}^\dagger
\!\!\!&=&\!\!\!\frac{1}{\sqrt{(E_{\bf k}+\omega_{\bf
k})^2+|\Delta|^2}}\left((E_{\bf k}+\omega_{\bf k})
\akdestroy{-k}+\Delta^\dagger \akdestroy{k}\right)\label{quasipart2}
\end{eqnarray}
which result in the diagonal Hamiltonian 

\begin{equation}
{\mathcal H}={\mathcal H}^0+\sum_{{\bf k}\neq 0} \omega_{\bf k}
\hat\alpha^\dagger_{\bf k}\hat\alpha^{\phantom\dagger}_{\bf k},
\end{equation}
where ${\mathcal H}^0$ contains the ground state contribution to the
energy and the excitations are given by the spectrum of frequencies
$\omega_{\bf k}=\sqrt{E_{\bf k}^2-|\Delta|^2}$.

Since there are no quasiparticles present in the ground state
$\big|gs\big>$ , which is what one would expect from an interacting
bosonic system at $T=0$, the ground state must satisfy
the condition \mbox{$\hat\alpha_{\bf k} \big|gs\big>=0$}. Substitution
of Eqs.~(\ref{quasipart1}) and~(\ref{quasipart2}) for the
quasiparticle operators result in the relation

\begin{eqnarray}
(E_{\bf k}+\omega_{\bf k})\akdestroy{k} \big|gs\big>&=&-\Delta
\akcreate{-k}\big|gs\big>.\label{prediffeq}
\end{eqnarray}
Because $\akdestroy{k}$ and $\akcreate{k}$ are canonically conjugate
variables there is no loss of generality in making the replacement
$\akdestroy{k}\rightarrow\partial/\partial \akcreate{k}$ \cite{dirac}.
This converts Eq.~(\ref{prediffeq}) into a simple differential equation
for the ground state with the solution

\begin{equation}
\big|gs\big>=\exp\left[\sum_{\bf k}\frac{-\Delta}{E_{\bf k}+\omega_{\bf k}}\akcreate{-k}
\akcreate{k}\right]\big|0\big>.\label{gs}
\end{equation}
To derive the many-body wavefunction we must now move from second to
first quantization.  The relationship which links the second quantized
ground state $\big|gs\big>$ with the first quantized wavefunction $\Psi_{CB}$ can be written for
an even number of noncondensed particles $2N$ as

\begin{equation}
\Psi_{CB}({\bf x}_1,{\bf x}_2,... {\bf
x}_{2N})=\big<0\big|\hat\varphi_a({\bf x}_{2N})...\hat\varphi_a({\bf x}_{\text
2})\hat\varphi_a({\bf x}_{\text 1})\big|gs\big>,
\end{equation}
where it should be noted that $\Psi_{CB}$ is the full-many body
wavefunction for the composite Bose particles. If we are able to
assume that $E_{\bf k}+\omega_{\bf k}\gg\Delta$ , an assumption which will remain
valid as long as we are not too close to resonance, we may truncate
the power expansion of the exponent in Eq.~(\ref{gs}). This results in the
composite boson wavefunction

\begin{equation}
\Psi_{CB}({\bf x}_1,{\bf x}_2,... {\bf
x}_{2N})={\mathcal S}(\psi_{12}\psi_{34}...\psi_{(2N-1)2N})\label{cp}
\end{equation}
comprised of a symmetrized product ${\mathcal S}$ of paired wavefunctions
 
\begin{equation}
\psi_{ij}=\sum_{\bf k}\frac{-\Delta}{E_{\bf k}+\omega_{\bf k}}e^{i{\bf k}\cdot({\bf
x}_i-{\bf x}_j)}.\label{iandj}
\end{equation}
If we were dealing with a system of fermions, equation~(\ref{cp}) would be
antisymmetrized and would result in a Pfaffian wavefunction
\cite{read}.  Here, because of the statistics of the particles,
we generate a bosonic analogue to this result.  The many-body wavefunction 
for the bare particles can now be extracted from the composite
wavefunction \cite{morinari} resulting in

\begin{equation}
\Psi_{MB}=\Psi_{CB}({\bf x}_1,{\bf x}_2...,{\bf
x}_{2N})\times \Psi_L ({\bf x}_1,{\bf x}_2...,{\bf
x}_{2N})\label{sp}\label{result}
\end{equation}
which is a product of the composite particle wavefunction of Eq.~(\ref{cp}) and the Laughlin wavefunction of Eq.~(\ref{laughlin}).

Equation~(\ref{result}) is the final result for the ground state
wavefunction of the resonant rotating Bose system.  This result has
important consequences for the generation of the FQHE within a
resonant atomic gas.  It would imply that an increase of interparticle
interactions by a resonant tuning of the interactions simultaneously
generates an increase in 2-particle correlations between composite
particles resulting in a modification to the ground state
wavefunction. As is clear from the form of Eq.~(\ref{cp}), for large
detuning from the resonance, corresponding to small pairing and
molecular field, the many-body wavefunction reduces to the Laughlin
wavefunction. As one moves nearer to the resonance, however, the
off-diagonal part of the self-energy matrix, $\Delta$, grows. This
results in an increasing modification of the many-body wavefunction
from the Laughlin wavefunction. The ability to tune a Feshbach
resonance therefore allows for the direct study of this crossover from
a Laughlin wavefunction to a paired wavefunction.\\ \indent A similar,
yet distinct, paired wavefunction as in Eq.~(\ref{result}) was found
for electronic FQHE systems \cite{haldane, read, greiter2}.  This has been
used to explain the previously unresolved even denominator filling
fractions which result from a pairing instability, such as the
observed incompressibility of the 5/2 filling.  In this case, a
straightforward generalization of the Laughlin wavefunction would
result in a symmetric wavefunction, violating the asymmetry of the
fermions.  However, the generation of an antisymmetric Pfaffian
wavefunction which multiplies the generalized Laughlin state allows
the overall ground state to be correctly antisymmetrized. For the
bosonic system we have treated, the overall wavefunction must remain
symmetric, so the corresponding paired wavefunction is symmetric in
comparison to the antisymmetric Pfaffian wavefunction.\\ \indent In
conclusion, because of the extreme diluteness of trapped atomic gases,
if strongly correlated effects such as the FQHE are to be observed in
these experiments, the interatomic interactions will most likely need
to be resonantly enhanced.  Feshbach resonances are a convenient
method for increasing the interactions, but the molecular processes
involved in such a resonant system force one to account for the
effects of atomic pairing. The growth of pair correlations among the
composite particles leads to a modification of the expected ground
state for a rapidly rotating Bose gas.  The new ground state
wavefunction, which is generated by the Feshbach resonance, exists as
a strongly correlated state unique to trapped Bose gases. The
tunability of the resonance opens the possibility for the direct study
of the crossover transition between the paired state and the Laughlin
state.\\\indent These results have important implications for the
production of the FQHE within atomic gases since the use of a Feshbach
resonance results in a state quite different from a rapidly rotating
gas where the interactions are quantified by only a large scattering
length. For instance, many of the observeable properties of the gas
may be modified such as the density profile for both atoms and
molecules and the nature of collective excitations. It should also be
noted that the crossover transition we have discussed is only a part
of a much more general crossover theory made accessible by the
tunability of a Feshbach resonance.  Although the methods presented
here are invalid close to the resonance, one could imagine extending
these ideas to describe the resonant system as one passes from a gas
of interacting rotating atoms, through the resonance, to a system of
tightly bound, rotating molecules.\\\indent Support is acknowledged
for S.B., J.M., and M.H. from the U.S. Department of Energy, Office of
Basic Energy Sciences via the Chemical Sciences, Geosciences and
Biosciences Division.

\end{document}